\documentclass[preprint2]{aastex}
\def\ltsima{$\; \buildrel < \over \sim \;$}
\def\simlt{\lower.5ex\hbox{\ltsima}}
\def\gtsima{$\; \buildrel > \over \sim \;$}
\def\simgt{\lower.5ex\hbox{\gtsima}}
\def\kms{{\rm\,km\,s^{-1}}}

\def\s{\ifmmode \widetilde \else \~\fi}
\def\={\overline}

\def\spose#1{\hbox to 0pt{#1\hss}}

\def\lta{\mathrel{\spose{\lower 3pt\hbox{$\mathchar"218$}}
     \raise 2.0pt\hbox{$\mathchar"13C$}}}
\def\gta{\mathrel{\spose{\lower 3pt\hbox{$\mathchar"218$}}
     \raise 2.0pt\hbox{$\mathchar"13E$}}}
\def\Dt{\spose{\raise 1.5ex\hbox{\hskip3pt$\mathchar"201$}}}	% upper case
\def\dt{\spose{\raise 1.0ex\hbox{\hskip2pt$\mathchar"201$}}}	% lower case

\def\dotsfill{\leaders\hbox to 1em{\hss.\hss}\hfill}

\slugcomment{To Appear in ApJ Letters}

\shorttitle{Lewis \& Ibata}
\shortauthors{Probing the Atmospheres of Microlensed Planets}
\begin{document}

\title{Probing the Atmospheres of Planets Orbiting Microlensed Stars via
Polarization Variability}


\author{Geraint F. Lewis}
\affil{Anglo-Australian Observatory, P.O. Box 296, Epping, NSW 1710, Australia}
\email{gfl@aaoepp.aao.gov.au}

\author{Rodrigo A. Ibata}
\affil{Max-Plank Institut f\"ur Astronomie, 
K\"onigstuhl 17, D--69117 Heidelberg, Germany}
\email{ribata@mpia-hd.mpg.de}



\begin{abstract}
We present a new method  to identify and probe planetary companions of
stars in the Galactic  Bulge and Magellanic Clouds using gravitational
microlensing.  While  spectroscopic studies  of these planets  is well
beyond  current   observational  techniques,  monitoring  polarization
fluctuations  during  high  magnification  events  induced  by  binary
microlensing  events  will  probe  the composition  of  the  planetary
atmospheres, an observation  which otherwise is currently unattainable
even for nearby planetary systems.
\end{abstract}

\keywords{Gravitational Lensing; Polarization; Planetary Systems}

\newcommand{\tauboo}{$\tau$ Bo\"{o}tis}

\section{Introduction}\label{introduction}
With recent  technological advances it  has become possible  to detect
extra-solar planets  via their  gravitational influence on  their host
stars.  Inducing  velocity shifts  of tens to  hundreds of  meters per
second,  accurate Doppler  measurements  have successfully  identified
$\sim30$  such  systems  (Mayor   \&  Queloz  1995;  Marcy  \&  Butler
1998)~\footnote{\tt
http://www.physics.sfsu.edu/$\sim$gmarcy/planetsearch/}.
While uncovering  the existence of planetary  systems, such techniques
do not probe directly the physical properties of the planets.  Rather,
the conditions of the planet are inferred from the orbital parameters.
However,  monitoring the  change  in stellar  brightness  as a  planet
transits  a   star  reveals  the  planetary   radius,  an  observation
successfully undertaken on HD~209458  (Charbonneau et al. 2000; Henry,
Marcy,  Butler, \&  Vogt  2000).  Furthermore,  a  novel approach  has
recently detected the Doppler-shifted  light reflected from a massive,
gas-giant planet orbiting  \tauboo.  Utilizing high resolution optical
spectroscopy,  the  variability  of  observed spectral  features  have
provided an  estimation of the planet's size,  assuming a Jupiter-like
albedo (Collier Cameron, Horne, Penny \& James 1999: CCHPJ99).

%While  breaking  new  ground   in  the  detection  of  planets,  these
These approaches are  currently only applicable to stars  within a few
tens of parsecs of  the Sun.  Gravitational microlensing, however, has
the potential to detect planetary systems at several kiloparsecs.  The
search for  microlensing by MACHOs within the  Galactic Halo, proposed
by Paczy\'{n}ski (1986)  and first detected by Alcock  et al.  (1992),
has become  an veritable  astronomical cottage industry,  with several
groups  monitoring stars  towards  various sources~\footnote{See  {\tt
http://wwwmacho.anu.edu.au/} for a description  of the MACHO group and
links to the other teams.}.
These studies  have proved to  be very successful,  identifying $>400$
candidates.  A  substantial fraction  of these events  display complex
light  curves,  characterized  by  rapid  and  multiple  fluctuations,
indicative of microlensing by a binary pair of compact objects (Alcock
et al. 2000). When the  mass ratio between the two microlensing bodies
is  very much  different from  unity, as  is the  case  with planetary
systems, the  result is less  dramatic with the resultant  light curve
exhibiting  the general  isolated body  bell-shaped profile,  with the
addition  of  a relatively  short  time  scale additional  fluctuation
(e.g. Mao \& Paczy\'{n}ski 1991; Gould \& Loeb 1992; Wambsganss 1997).
Several  halo  microlensing light  curves  are  seen  to possess  such
fluctuations, indicative of the existence of planetary mass companions
of the  MACHO bodies (e.g.  Bennett  et al.  1999; Rhie  et al.  1999;
Alcock  et al.   2000), although  the interpretation  of  these events
still remains controversial (Albrow et al. 2000).

These  approaches  focus  upon  the perturbative  effects  induced  by
planets  orbiting the  MACHO  lensing objects  induce  on the  caustic
structure.  Here we consider instead planets orbiting the microlensing
sources,  namely stars in  the LMC,  SMC and  the Galactic  Bulge, and
examine how  large magnifications, induced when  the planetary systems
is swept by  a fold caustic, can probe the  physical properties of the
planet.

During the final  stages of the preparation of  this paper, the recent
work of Graff  \& Gaudi (2000) was brought to  our attention. They too
consider  the discovery  of  planets  in the  Galactic  Bulge via  the
identification  of  additional photometric  features  due to  caustics
sweeping  across  the  planetary  system.   Their  results  and  those
presented  here for  the  photometric light  curves  are in  excellent
agreement.

\section{Gravitational Microlensing}\label{microlensing}
The mathematical basis to the simple microlensing seen in the Galactic
halo has  been presented elsewhere,  and will not be  reproduced here.
The  reader  is  directed  towards  the excellent  review  article  by
Paczy\'{n}ski (1996). The important  scale length, the Einstein radius
in the plane of the lensing masses, is defined to be
\begin{equation}
R_E = \sqrt{ \frac{4 G M}{c^2}\frac{D_{ls}D_{ol}}{D_{os}} }\,  = 
8.09 \left[\frac{M}{M_\odot}
\frac{D_{os}}{8kpc}
\left(1-d\right)d\right]^{\frac{1}{2}} AU\, ,
\label{einstin}
\end{equation}
where $D_{os}$,  $D_{ls}$ and $D_{ol}$  are the distances  between the
source   ($s$),  lens   ($l$)   and  observer   ($o$),   and  $d$   is
$D_{ol}/D_{os}$.   For a solar  mass star  located midway  between the
Earth  and  the  Galactic  Bulge ($D_{os}=8kpc$),  $R_E=4.05AU$.   The
corresponding  value for  microlensing  of sources  in the  Magellanic
Clouds ($D_{os}=55kpc$) is $10.62AU$.

\subsection{High Amplification Events}\label{hae}
A  single, isolated  mass produces  a very  simple  point-like caustic
structure, and for  a source to be significantly  magnified it must be
extremely well aligned with  this caustic feature.  When, however, the
lens consists of  a binary pair of masses, the  combination of the two
lensing  potentials unfolds  the point-like  caustic structure  into a
more  complicated,  extended form  (Schneider  \&  Weiss 1986).   This
caustic structure  is dominated by `fold catastrophes',  for which the
magnification is singular along a line,  as opposed to a point for the
isolated mass case.   Due to their extended nature,  fold caustics are
powerful probes of emission structure in a source, and can potentially
unravel  the  complex  structure  of  the central  regions  of  active
galaxies on a scale far smaller than the resolution of current optical
telescopes  (Wambsganss \&  Paczy\'{n}ski 1991;  Lewis \&  Belle 1998;
Belle  \& Lewis  2000).  They  can  also reveal  important details  of
stellar  surfaces,   such  as  the   extent  of  limb   darkening  and
polarization  distribution  (e.g.   Simmons,  Willis \&  Newsam  1995;
Simmons, Newsam \& Willis 1995).

The  magnification induced  as a  point source  is crossed  by  a fold
caustic is given by
\begin{equation}
\mu = \mu_o + \frac{k}{\sqrt{x - x_c}} H\left(x-x_c\right)\,,
\label{caustic}
\end{equation}
where  $x-x_c$ is  the distance  from the  source to  the  caustic and
$H(x)$  is the  Heaviside step  function (Schneider,  Ehlers  \& Falco
1992).  The  `strength' of the  caustic is denoted by  the flux-factor
$k$ and  typically has  a value  of unity (Kayser  \& Witt  1989; Witt
1990; Witt, Kayser \& Refsdal 1993).

\subsection{Source Model}\label{source}
Of the sample of extra-solar planets known currently, all are at least
a substantial fraction of a Jupiter mass.  All are seen to orbit close
to their  host star, with the  majority of semi-major  axes being less
than   $\sim1AU$.   Such  a   distribution  may   be  the   result  of
observational selection  effects, because  it is these  massive, close
orbiting planets  that would produce the largest  Doppler reflex.  For
the purposes of  this study, the source is taken to  be similar to the
recently identified \tauboo\ planetary system (CCHPJ99).

\tauboo, an F7V  star, has a mass of 1.2$M_\odot$  and a luminosity of
3.5$L_\odot$.  The Doppler reflex  motion of amplitude $152\kms$ and a
period  of  $3.3$~days  suggests  a  planetary  companion  of  8$M_J$,
orbiting  at $4.62\times10^{-2}AU$  (neglecting  inclination effects).
With such a proximity to its  host star, the planet can be expected to
reflect  a nonnegligible fraction  of light  towards an  observer, and
CCHPJ99 measure $\epsilon\sim2\times10^{-4}$ of the flux from \tauboo\
is light  reflected from the planet  (but see also  Charbonneau et al.
1999 \& Burrows  et al. 2000).  With this, the  planet has an inferred
radius  of  1.8$R_J$  (for  CCHP99's  preferred  albedo).   While  the
characteristics  of the planet  orbiting \tauboo\  may be  extreme, it
will serve to illustrate the details of the microlensing model.

\subsection{Maximum Magnifications}\label{maxmag}
The magnification  of a  point-like source is  singular at  a caustic.
Such singularities  are, however, integrable for  any extended source,
resulting in a finite magnification.  The maximum by which a source is
enhanced as a caustic sweeps across it is given by (Chang 1984);
\begin{equation}
\mu_{max}\sim \frac{f k }{\sqrt{R_s}}
\label{max}
\end{equation}
where $R_s$  is the  effective radius  of the source  in units  of the
Einstein radius,  $k$ is the  flux-factor (Equation~\ref{caustic}) and
$f$ is  a form factor which  is dependent upon  the surface brightness
distribution of the source [$f$ is typically $\sim1.5$, being 1.39 for
a uniform disk  (Kayser \& Witt 1989)].  For  a given caustic, smaller
sources can be enhanced to a greater degree.  Sources with a radius of
$1R_\odot$, located  in the Galactic  Bulge, are subject to  a maximum
magnification  of  $\mu_{max}\sim70$,  while  for  a  body  of  radius
$\sim1.8R_J$ (equivalent  to the planet orbiting  \tauboo), this value
increases to  $\sim161$.  Here, the  flux-factor is assumed to  be 1.2
(Wook, Chang \&  Kim 1998).  Taking account of  the relative distances
the  corresponding   values  of  magnification  for   sources  in  the
Magellanic Clouds are $\sim62\%$ larger.

When observing a  microlensing event the light detected  is a blend of
the star and the reflected  light from the planet.  In this situation,
the maximum fluctuation  induced in a light curve  as a caustic sweeps
across a planet is given by
\begin{equation}
\delta   M_{max}   \sim   2.5   \log\left(  1   +   \epsilon   \frac{k
f}{\sqrt{R_s}}\right)\, .
\label{maxmagnitude}
\end{equation}
Employing  the parameters  for  the \tauboo\  system described  above,
$\delta M_{max} \sim 0.035$. The corresponding value for a planetary 
system located in the LMC/SMC is $\delta M_{max} \sim 0.055$

\subsection{Microlensing Light Curves}\label{events}
\begin{figure*}
\centerline{ 
\includegraphics[angle=270,width=3.2in]{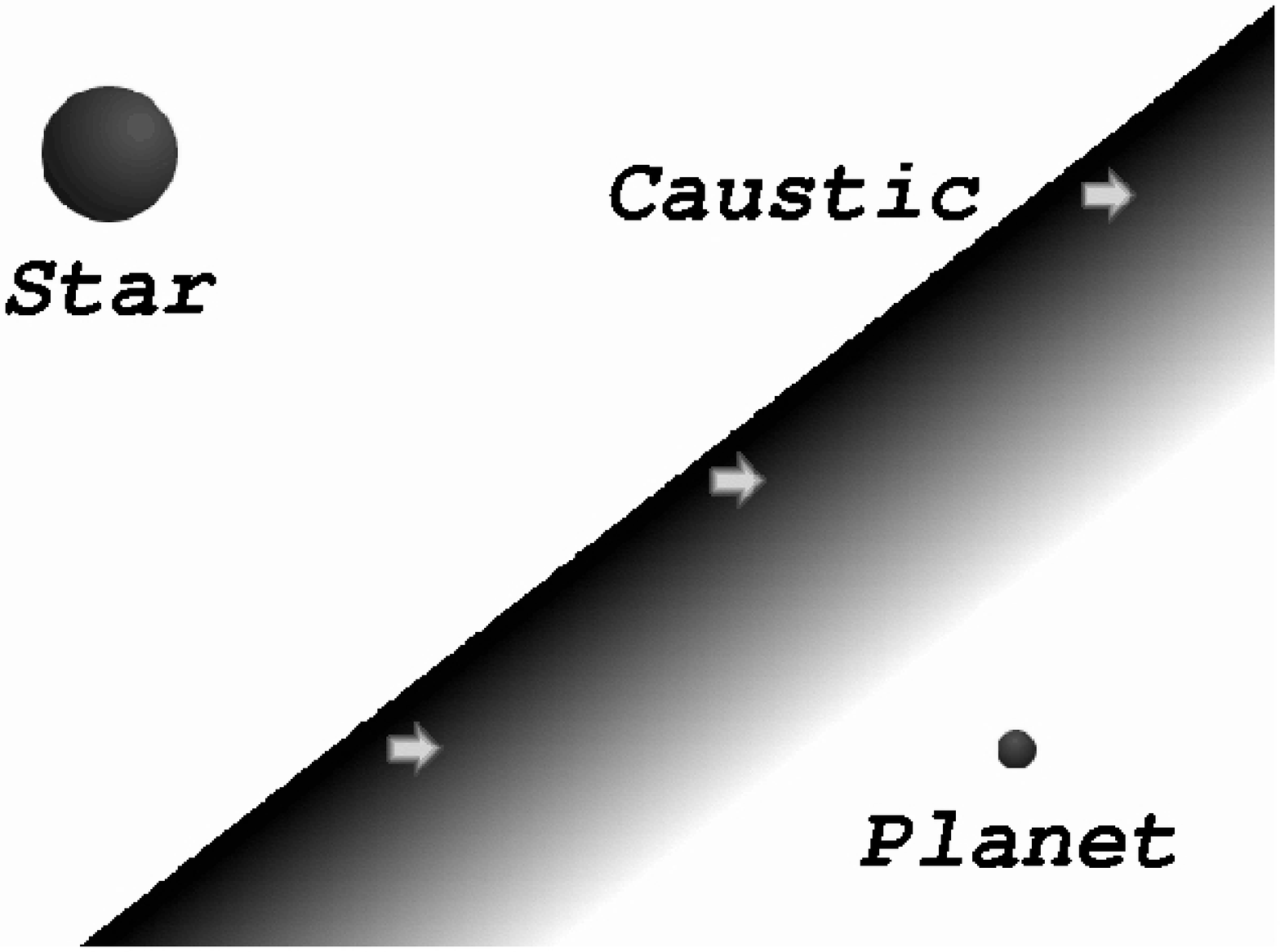}}
\caption[]{Schematic representation of the microlensing configuration.
The  grey shaded  denotes  the distribution  of  magnification at  the
caustic edge, while  the arrows denote its direction  of motion. Here,
the caustic has  swept across the star, which now lies  in a region of
lower magnification, and is about to cross the planet.}
\label{fig1}
\end{figure*}

Combining the microlensing and planetary model, a schematic picture of
the   microlensing   configuration    of   interest   can   be   drawn
(Figure~\ref{fig1}). It is assumed that  the star is observed near the
end of a binary microlensing event, recognized through its distinctive
light curve, and that the star has moved out of the caustic structure.
Here it is  magnified by $\mu_o\sim1$.  The caustic  has, however, yet
to fully leave  the stellar system and has yet  to cross the planetary
companion.

\begin{figure*}
\centerline{ 
\includegraphics[angle=270,width=3.2in]{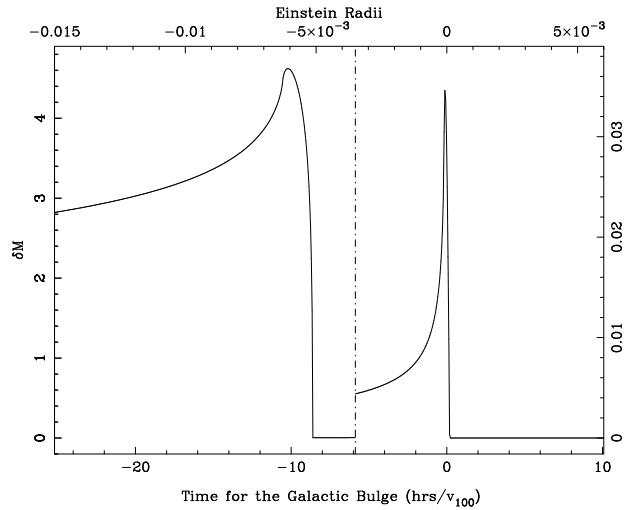}}
\caption[]{The light  curve produced by a caustic  sweeping across the
planetary system.   The lower abscissa represents the  time (in hours)
for a microlensing event in  the Galactic Bulge while the upper values
correspond to Einstein  radii for a solar mass  star.  The ordinate is
change in  magnitudes for the stellar event  (left-hand) and planetary
crossing (right-hand). Note the change of scales for each event.  }
\label{fig2}
\end{figure*}

Sweeping a caustic across the model \tauboo\ system produces the light
curve presented in Figure~\ref{fig2}.   The upper abscissa is in units
of  Einstein radii  for a  Solar mass  star, and  the ordinate  is the
change  in brightness of  the microlensed  source.  The  caustic moves
from left  to right, first crossing  the primary star  and producing a
fluctuation  of more than  four magnitudes.   After this,  the caustic
moves  on over  the planet  at  $x=0$.  Due  to its  smaller size,  it
undergoes a  sharper, stronger magnification,  although as it  is much
fainter  than its  host star,  it produces  a much  smaller  change in
magnitude
~\footnote{The lensing situation described  above is, of course, fully
time  reversible.  Given an  initial alert  at the  start of  a binary
event, however,  later high  magnification crossings can  be predicted
and detailed observations scheduled.}.

The conversion of the abscissa in Figure~\ref{fig2} into observed time
is given by:
\begin{equation}
t_{E} = \frac{R_E}{v_m},
\label{time_vel}
\end{equation}
where  $v_m$ is  the transverse  velocity of  the microlenses.   For a
source  in the  Galactic Bulge,  with the  microlenses  located midway
between  the  source  and  the  observer, and  assuming  a  transverse
velocity  difference of $100v_{100}\kms$,  $t_E=70 v_{100}^{-1}$~days;
this  has been  used to  normalize Figure~2  and represents  the lower
abscissa coordinates.  For the  Magellanic Clouds, where we can assume
a  higher velocity  of  $200v_{200}\kms$, the  corresponding value  is
$t_E=92 v_{200}^{-1}$~days.  Following Figure~2, as the caustic sweeps
across the planetary  system it first crosses the  host star, strongly
magnifying it for several hours, with the apparent brightness dropping
rapidly after the caustic has completely transited the star. The light
curve is then  quiescent as the caustic travels  the distance from the
star to  the planet.   The semi-major axis  of the \tauboo\  planet is
inferred to be 0.0462AU; projecting into the plane of the microlensing
mass,  this corresponds  to $5.7\times10^{-3}  R_E$ $(2.2\times10^{-3}
R_E)$ for the  Galactic Bulge (Magellanic Clouds), and  the time taken
for the caustic to  sweep across this distance is 9.6$v_{100}^{-1}$hrs
(4.8$v_{200}^{-1}$hrs.  After this  time scale,  the region  of strong
magnification associated with the  caustic reaches the planet, and the
planet  undergoes  strong  magnification $(\simgt50)$  for  1-2~hours.
Again,  once  the caustic  has  completely  transited  the planet  its
apparent  brightness  falls   rapidly.   Given  the  relatively  short
duration of  such a planetary microlensing `event',  orbital motion of
the planet has been neglected.

\section{Photometric and Polarmetric Monitoring}\label{observations}
Graff  \&  Gaudi (2000)  have  considered  the  identification of  the
photometric  variability  induced  by  the  presence  of  a  planetary
companion of a microlensed source in the Galactic Bulge, demonstrating
that  monitoring on  a 10-m  class telescope,  with  5-minute temporal
resolution,  would  provide a  detection  with  a $S/N\sim20$.   While
revealing  the planet's  presence, however,  determining  its physical
properties is much more difficult, as any spectroscopic study would be
swamped by  the flux from the  host star and  observations to separate
the two  components would require  integration times much  longer than
the time scale of any variability.

Recently, Seager, Whitney \& Sasslov (2000) investigated the potential
nature  of  the  planetary  atmospheres,  focusing  on  the  particles
responsible for reflecting the light from the host star. This material
also  acts  to  polarize  the  reflected  flux,  with  the  degree  of
polarization  being  highly  dependent  upon  the  particle  size  and
composition,  and  hence the  physical  conditions,  in the  planetary
atmosphere.  In practice, however,  the degree of polarization is very
small, ranging up  to $\sim 10^{-5}$, dependent on  the model geometry
and  the dominant particle  size in  the planetary  atmosphere.  These
polarization values  are small due  to the dominating presence  of the
unpolarized  flux  from the  host  star.  Flat-fielding  uncertainties
coupled  with  inevitable  variable   seeing  limit  the  accuracy  of
polarization measurements,  and as  Seager, Whitney \&  Sasslov (2000)
conclude,  the expected  values  are well  below  the capabilities  of
current observational techniques, even for nearby systems.

Can  gravitational  microlensing  boost  the polarization  to  a  more
observable  level?  Considering  a  planetary system  which induces  a
fractional polarization  of $P_o$, the value observed  when the planet
is magnified by a factor $\mu$ is
\begin{equation}
P_\mu = \mu \frac{ 1 + \epsilon }{ 1 + \mu \epsilon} P_o \ .
\label{fractional polarization}
\end{equation}
Using the  maximum magnification of planets described  in Section 2.3,
the  maximum observed  fractional polarization  during  a microlensing
event is $\sim156P_o$ for a \tauboo-like system in the Galactic Bulge.
The corresponding number for the Magellanic clouds is $\sim247P_o$ and
so the apparent flux from  the systems is substantially more polarized
during a planetary microlensing event.

Scaling the Seager, Whitney  \& Sasslov (2000) fractional polarization
estimates  from their 51Peg  model to  \tauboo\ the  maximum intrinsic
V-band  polarization  will  be  $P_o=1.5 \times  10^{-5}$.   Thus  the
magnified  fractional polarization will  be $\sim0.23\%$  for Galactic
Bulge sources,  and $\sim0.37\%$ when microlensing sources  are in the
Magellanic Clouds.  While these  values focus upon the `best-case', an
examination of  Fig 2. reveals  that the planet will  be significantly
magnified for  an extended  period (by $\simgt50$  for an hour  in the
Galactic  Bulge) and  so deeper  limits can  be reached  with extended
integration times.

It is  currently routine to  measure polarization at the  0.1\% level,
and polarimeters  are now available  on large telescopes,  making such
measurements  possible  even  for   faint  sources.   With  the  FORS1
polarimeter instrument on ESO's Very Large Telescope, a $\sim 400$~sec
exposure at  two position angles is sufficient  to detect $\sim0.23\%$
fractional  polarization in  the V-band  at the  $3\sigma$ level  on a
${\rm V=18}$  source (corresponding to  the the apparent  magnitude of
\tauboo\ at the distance of the Galactic Bulge). Including the effects
of extinction, assuming an $A_v\sim 1.25$, the integration needs to be
increased to  $\sim700$~sec to  achieve the same  detection.  Figure~2
reveals that, for the Galactic Bulge, the magnification exceeds 65 for
$\sim35$~mins, and hence the polarization exceeds $\sim0.1\%$ over the
same period. As there will be  a distribution in the luminosity of the
sources under consideration, the  observing strategy can be `tuned' to
provide  significant detections/limits.   For \tauboo-type  systems in
the Magellanic Clouds  (${\rm V \sim 22}$), however,  the detection of
the   expected   polarization    appears   unfeasible   with   current
instrumentation, as  the required integration  times become comparable
to the planet crossing time.

\section{Conclusions}\label{conclusions}
This paper  has considered the identification of  planets orbiting the
sources  (rather  than  the  lenses) monitored  by  the  gravitational
microlensing experiments,  namely stars in the Galactic  Bulge and the
Magellanic Clouds.   In this case it  is the light  reflected from the
planetary  surface  that   undergoes  significant  magnification  when
crossed by  a caustic.   A close-in extrasolar  giant planet,  such as
that  identified  orbiting  \tauboo,  would induce  a  fluctuation  of
$\delta M_{max} \sim 0.035 (0.055)$ into a microlensing light curve of
a   source  in   the  Galactic   Bulge  (Magellanic   Clouds),  easily
identifiable a few hours before  or after maximum magnification with a
monitoring  program  on  a  10-m  class  telescope;  such  photometric
variations  will  be  observable   in  both  the  Galactic  Bulge  and
Magellanic  Clouds.  It  should  be noted  that,  with the  photometry
limits  available   with  current  telescopes,  the   latter  case  is
restricted   to  luminous   stars,  such   as  evolved   giants.   The
determination of  the physical properties  of the planet  is, however,
more difficult due  to the relatively short duration  of the planetary
event and the overwhelming flux from the host star.

The light reflected from close-in extrasolar giant planets is expected
to be  polarized, due  to the presence  of particles in  the planetary
atmosphere. The degree of  polarization is strongly dependent upon the
composition and particle size and hence the physical conditions of the
atmosphere.  If unmagnified, this  induced polarization is small, well
below the detection capabilities of modern telescopes.

As  a  gravitational  microlensing  caustic crosses  the  planet,  the
expected degree of  induced polarization can be brought  to levels now
detectable with  10-m class  telescopes.  Seager, Whitney  \& Sasselov
(2000), however,  concede that significant  uncertainties still remain
in  their  modeling of  the  fractional  polarization  induced by  the
presence  of  a  planetary  companion.  Additional  effects,  such  as
multiple  scattering and  blending  in crowded  fields,  which act  to
reduce the polarization to even  smaller values, may take the proposed
probe of planetary atmospheres out of reach with current observational
techniques.  Nevertheless, the huge gain  in contrast in the flux of a
planet compared  to that  of its host  during a caustic  crossing will
likely  make this  an attractive  strategy for  future investigations.
While such observations of  the Magellanic Clouds are currently beyond
current techniques, the  polarmetric monitoring of microlensing events
in the Galactic Bulge may reveal more clues to the physical conditions
of close-in extrasolar giant planets than studies of planetary systems
much nearer to the Earth.

\section*{Acknowledgments}
We thank  E. Corbett  and J. Bailey  for very useful  discussions. The
anonymous  referee is  thanked for  providing useful  comments  on the
original  version of  this  manuscript.

\newpage

\label{lastpage}

\end{document}